\documentclass{article}
\usepackage{latexsym}
\usepackage{color,cite,graphicx,setspace}
\usepackage{latexsym,amssymb,epsf,subfigure} 

\newcommand{\bfc}{{\bf c}}
\newcommand{\bfE}{{\bf E}}
\newcommand{\bfJ}{{\bf J}}
\newcommand{\bfn}{{\bf n}}
\newcommand{\bfp}{{\bf p}}
\newcommand{\bfP}{{\bf P}}
\newcommand{\bfx}{{\bf x}}

\newcommand{\bfsigma}{{\mbox{\boldmath $\sigma$}}}

\newcommand{\calE}{{\cal E}}
\newcommand{\calJ}{{\cal J}}
\newcommand{\calL}{{\cal L}}
\newcommand{\calN}{{\cal N}}
\newcommand{\calP}{{\cal P}}
\newcommand{\calZ}{{\cal Z}}
\newcommand{\bfxt}{\left(\bfx,t\right)}
\newcommand{\bfxtp}{\left(\bfx+\bfc_i,t\right)}

\begin{document}
\doublespacing
\title{
   Three dimensional hydrodynamic lattice-gas simulations of domain growth
  and self-assembly in binary immiscible and ternary
   amphiphilic fluids.
}
\author{
  Peter J. Love\\
  {\footnotesize Theoretical Physics, Department of Physics, University of Oxford,}\\
  {\footnotesize 1 Keble Road, Oxford, OX1 3NP, UK}\\
  {\footnotesize{\tt love@thphys.ox.ac.uk}}\\[0.3cm]
  Peter V. Coveney,\\
  {\footnotesize Centre for Computational Science,
    Queen Mary and Westfield College,}\\
  {\footnotesize University of London, Mile End Road,}\\
  {\footnotesize London E1 4NS, U.K.}\\
  {\footnotesize{\tt p.v.coveney@qmw.ac.uk}}\\
  and\\[0.3cm]
  Bruce M. Boghosian,\\
  {\footnotesize Department of Mathematics,
    Tufts University,}\\
  {\footnotesize 211 Bromfield-Pearson Hall, Medford, Massachusetts,
    02155, U.S.A.}\\
  {\footnotesize{\tt bruce.boghosian@tufts.edu}}
}

\maketitle
\newpage
\begin{abstract}
We simulate the dynamics of phase assembly in binary immiscible fluids
and ternary microemulsions using a three-dimensional hydrodynamic
lattice gas approach. For critical spinodal decomposition we perform
the scaling analysis in reduced variables introduced by
~\cite{bib:jury,bib:kendon}. We find a late-stage scaling exponent
consistent with the $R \sim t^{2/3}$ inertial regime. However, as
observed with the previous lattice-gas model of Appert {\it et
al.}~\cite{bib:appert} our data does not fall in the same range of
reduced length and time as that of Kendon {\it et al.}~\cite{bib:jury,bib:kendon}. For off-critical binary spinodal decomposition we observe a reduction
of the effective exponent with the volume fraction of the minority
phase. However, the $n=\frac{1}{3}$ Lifshitz-Slyzov-Wagner droplet
coalescence exponent is not observed.  Adding a sufficient number of surfactant particles to
a critical quench of binary immiscible fluids produces a ternary bicontinuous
microemulsion. We observe a change in scaling behaviour from algebraic
to logarithmic growth for amphiphilic fluids in which the domain growth is not
arrested. For formation of a microemulsion where the domain growth is
halted we find a stretched exponential growth law provides the best
fit to the data.
\end{abstract}

\section{Introduction}
The study of phase ordering kinetics has become a testbed for new
complex fluid simulation methods. Despite intensive analysis by
many methods, it remains  a field in which many interesting and
fundamental questions go unanswered. Constructing a model which correctly includes
hydrodynamics but which is computationally simple enough to reach
late times is a major theoretical challenge. New mesoscale
models such as lattice gas (LG)~\cite{bib:rk,bib:appert}, lattice Boltzmann
(LB)~\cite{bib:alexchen,bib:kendon}, dissipative particle dynamics (DPD)~\cite{bib:jury}, and Boltzmann-Vlasov~\cite{bib:basleb}
treatments have been crucial in increasing our understanding of these
systems over the past decade. Such methods require significantly
smaller computational resources than earlier molecular dynamics
(MD) and Cahn-Hilliard approaches, and can therefore access more
easily the late time regime in which hydrodynamics plays an important role.

A much studied system which exhibits hydrodynamic influences
on phase segregation is a $1:1$ mixture of immiscible fluids quenched
into the two phase region of its phase diagram (a so-called {\it
  critical} quench). Such a system undergoes spinodal decomposition,
where initially bicontinuous domain structure coarsens by surface-tension driven flows. Spinodal decomposition elicits much interest
because of its fundamental and industrial importance. For example, the
mechanical properties of alloys depend on the dynamics of the phase separation
process. Despite extensive theoretical
~\cite{bib:bray,bib:siggia,bib:kum1}, numerical ~\cite{bib:chak,bib:shinoono,bib:kogak,bib:pd,bib:fv,bib:lookwu,bib:ma,bib:larad1,bib:rk,bib:appert,bib:jury,bib:alexchen,bib:kendon,bib:basleb}
and experimental investigation ~\cite{bib:hasjinhas,bib:gue1}
some doubt remains about the true asymptotic late time growth
behaviour of these systems. 

The dynamics of phase segregation in binary immiscible fluids are also
dependent on the composition of the mixture. Mixtures which do not have
a $1:1$ ratio of the species (so-called {\it off-critical} quenches)
are much less studied than their critical counterparts. As the
quantity of the minority phase decreases, the domain structure ceases to
be bicontinuous, so that nucleation and coalescence of droplets
dominate the coarsening mechanism. 

The addition of surfactant to a system of binary immiscible fluids
radically alters the equilibrium properties and growth dynamics of
such mixtures~\cite{bib:gs2}. In particular, the adsorption of
surfactant (amphiphilic) molecules at oil-water interfaces leads to a
dramatic reduction in the interfacial tension. This property is
the origin of much industrial interest in surfactant systems.  
Only recently have computational techniques and sufficiently powerful
parallel platforms  become available which permit numerical simulation
of the hydrodynamic behaviour of amphiphilic fluids in three dimensions. Our hydrodynamic lattice gas model has
recently been implemented in three dimensions, and its ability to
reproduce many important phenomenological features of amphiphilic
systems confirmed~\cite{bib:bcp}. In the present paper 
we demonstrate this model's ability to capture quantitative features
of the dynamics of binary immiscible and ternary amphiphilic fluids.

Continuum approaches regard spinodal decomposition as a
solution of the equations of motion for two phase flow. These are
given by the Navier-Stokes equation within each phase, and by boundary
conditions at the interfaces. The incompressible Navier-Stokes
equations are, for each fluid phase:
\begin{equation}\label{eq:NS}
\rho \left(\frac {\partial {\bf v}} {\partial t} + {\bf v} \cdot
\nabla  {\bf v} \right) = \eta  \nabla^{2}
{\bf v} - \nabla p 
\end{equation}
\begin{equation}
\nabla \cdot {\bf v}=0
\end{equation}
where ${\bf v}$ is the fluid velocity, $\eta$ is the shear
viscosity and $p$ is the scalar pressure. 
The first boundary condition is that the velocity of the two phases must
match that of the interface:
\begin{equation}\label{eq:bc1}
{\bf u}_1 \cdot {\bf n} = {\bf u}_2 \cdot
{\bf n} =  {\bf u}_{int} \cdot {\bf n}
\end{equation}
where ${\bf n}$ is the interface normal and ${\bf u}_1$,
${\bf u}_1$ and ${\bf u}_{int}$ are respectively the velocities of
phase $1$ and $2$ at the interface, and the velocity of the interface
itself.
The second boundary condition requires that the stress difference at
the interface is balanced by the interfacial tension:
\begin{equation}\label{eq:surf_tens}
\underline{\bf T}_1 \cdot {\bf n}-\underline{\bf T}_2 \cdot
{\bf n} = \sigma \left( \frac{1}{R_a} +\frac{1}{R_b} \right){\bf n}
\end{equation}
where $\underline{\bf T}_1$ and $\underline{\bf T}_2$ denote the
stress tensor in phases $1$ and $2$, $R_a$ and $R_b$ are the two
principal radii of the interfaces and $\sigma$ is the interfacial tension.

Clearly, a numerical integration scheme for the above equations 
would be of enormous computational complexity. In order to obtain a numerical
model which captures the essential physics in a computationally
tractable way, an alternative approach is required. Remaining with a
phenomenological description of the fluid system, it is convenient
to define a scalar {\it order parameter} $\psi$:
\begin{equation}\label{eq:orderparam}
\psi({\bf x})=\rho_{o}({\bf x})-\rho_{w}({\bf x})
\end{equation}
where $\rho_{o}({\bf x})$ and $\rho_{w}({\bf x})$ are the densities of
oil and water at position ${\bf x}$. One may then write a
Cahn-Hilliard equation for the evolution of the order parameter:
\begin{equation}\label{eq:cahnhilliard}
\frac {\partial \psi} {\partial t} + {\bf v} \cdot
\nabla \psi = \lambda \nabla^{2} \mu + A\xi,
\end{equation}
and,
\begin{equation}\label{eq:functderiv}
\mu=\frac{\delta F}{\delta \psi},
\end{equation} 
and 
\begin{equation}\label{eq:freeng}
F=\int d^{d} {\bf x} \left[ \frac{1}{2} \psi^4 - \psi^2 + \sigma
  (\nabla \psi)^2 \right]. 
\end{equation}
This equation is coupled to the Navier-Stokes equations through a forcing term proportional to the gradient of
the chemical potential:
\begin{equation}\label{eq:NS2}
\rho \left(\frac {\partial {\bf v}} {\partial t} + {\bf v} \cdot
\nabla  {\bf v} \right) = \eta  {\bf \nabla}^{2}
{\bf v} - {\bf \nabla} p - \psi {\bf \nabla} \mu + A\nu.
\end{equation}
where $A\xi$ and $A{\bf \nu}$ are gaussian noise fields satisfying
an appropriate fluctuation-dissipation theorem ~\cite{bib:fv}.

%new paragraph: you've described model H, now describe how people
%solve it.

The equations~(\ref{eq:orderparam}~-~\ref{eq:NS2}) define a model for
 immiscible fluid dynamics commonly referred to as Model H.
This model enables one to obtain the scaling regimes for
critical spinodal decomposition by dimensional analysis. The {\it viscous} regime
is obtained when one may neglect the inertial terms on the left hand
side of equation~(\ref{eq:NS2}); then, balancing the forcing terms by the
viscous terms one obtains:
\begin{equation}
R(t) \sim \frac{\sigma}{\eta} t
\end{equation}
The {\it inertial} regime is obtained by balancing the forcing terms
$\psi {\bf \nabla} \mu$ by the inertial terms in eqn ~(\ref{eq:NS}):
\begin{equation}
R(t) \sim (\frac{\sigma}{\rho})^{1/3} t^{2/3}
\end{equation}
Equating these two regimes implies that the crossover from viscous to
inertial scaling occurs at $R \sim \eta^2 / \sigma\rho$. However, no
three-dimensional simulation method so far developed can reach a sufficient range of
length and timescales to observe both viscous and inertial regimes in
a single simulation. The work of Jury {\it et al.}~\cite{bib:jury} and
Kendon {\it et al.}~\cite{bib:kendon} overcomes
this difficulty by introducing scaling variables $L_0$ and $T_0$ which
enable data from separate simulations to be combined.

If the only physical effects involved in critical spinodal decomposition are
capillary forces, viscous dissipation and fluid inertia, then the
parameters governing domain growth are the surface tension
$\sigma$, fluid mass density $\rho$, and viscosity $\eta$. 
As emphasised by Jury {\it et al.}~\cite{bib:jury}, only one length,
${L_0}={\eta^2}/\rho\sigma$, and one time
$T_{0}={\eta^3}/\rho{\sigma^2}$ can be constructed from these
parameters. Data sets (L,T) from any model of spinodal decomposition
can be expressed in units of reduced length:
\begin{equation}\label{eq:red_len_defn}
l=L/L_0,
\end{equation}
and time,
\begin{equation}\label{eq:red_time_defn}
t=T/T_0.
\end{equation}
If all other physics is excluded from late-stage growth,
then the dynamical scaling hypothesis~\cite{bib:siggia} states that: 
\begin{equation}
l(t) \sim a+f(t),
\end{equation}
where $l(t)$ is the domain size expressed in reduced length units and $a$ is a non-universal constant which allows for a
period of growth in which molecular diffusion leads to the formation
of sharp interfaces. The function $f(t)$ should then approach a
universal form, identical for all $50:50$ incompressible binary fluid
mixtures following a deep quench.

The Model H equations have been integrated numerically in three
dimensions both with $A \neq 0$
(a quench to finite temperature), and with $A=0$  (a quench to zero
temperature) by a number of integration schemes ~\cite{bib:kogak,bib:lookwu}.
The free energy defined in eqn~(\ref{eq:freeng}) is also the basis for
the lattice-Boltzmann method of Kendon {\it et al.}~\cite{bib:kendon}. Although the
lattice-Boltzmann method has a physical motivation which replaces
velocity and density fields by single particle distribution functions,
it is essentially equivalent to direct numerical integration of Model H
such as that performed in ~\cite{bib:kogak,bib:lookwu} (i.e. it
should be regarded as a finite-difference scheme for solving these
equations in the absence of noise).  
The work of ~\cite{bib:kendon,bib:kogak,bib:lookwu} is therefore a confirmation of the scaling laws
derived above. The contribution of Kendon {\it et al.}~\cite{bib:kendon},
in whose work both growth regimes were clearly seen for
the first time, shows that the simple arguments given above are
incorrect in the crossover regime, and that the 
crossover in Model H extends over  $10^2 \geq t \geq 10^6$ in reduced time
units. It should be noted that this crossover was previously believed
from simple scaling arguments to occur at $t=1$.

Recently there has been further theoretical work by Grant and Elder:
those authors suggest ~\cite{bib:ge}
that growth in the inertial regime implies a Reynolds
number that increases without limit. This would
eventually lead to turbulent remixing of the fluids; the requirement
that the Reynolds number be self-limiting (i.e. that the critical
Reynolds number is not reached and therefore that turbulent remixing
does not occur) in the asymptotic regime implies a growth exponent of
$\leq 1/2$. There is at present no numerical evidence for the $n \leq
1/2$, and recent theoretical challenges to this idea have been
made. Novik and Coveney point out in~\cite{bib:novik1} that the relative strength of the interface
(characterised by the Weber number) must also be taken into account. A
small Weber number could delay the onset of turbulent remixing indefinitely.

The hydrodynamic lattice gas used in the present paper does
not rest on the macroscopic free energy functional proposed in eqn
~(\ref{eq:freeng}). The model, which is particulate in nature, is
described in detail below, and has a theoretical justification from
a `bottom up' perspective. In a bottom-up view the lattice-gas technique
may be regarded as a simplification of the molecular dynamics of a
binary fluid, abstracting the key microscopic properties in a
fictitious microworld.  Recent work has derived a microscopic basis
for the Rothman-Keller model of binary immiscible fluids~\cite{bib:rk,bib:micrork}. However, no
systematic method exists for coarse-graining a real molecular dynamics
description of a system to the lattice-gas model used here
(although such a systematic method does now exist for the DPD
algorithm~\cite{bib:flekcov}). 

In our model equations~(\ref{eq:NS}-\ref{eq:surf_tens}) are emergent macroscopic properties. The single phase FHP
lattice gas is known to
satisfy eqn.~(\ref{eq:NS}), and far from interfaces this behaviour is
reproduced in our model~\cite{bib:wolfram,bib:fchc}. In
Section~\ref{sec:st} we demonstrate that our model has
realistic surface tension behaviour at interfaces.

The purpose of the present paper is to investigate the dynamics of domain
growth of both binary immiscible and ternary amphiphilic phases in our
model. Section~\ref{sec:model} contains a brief description of our
model, while Section~\ref{sec:measure} contains a description of the
quantitative measurements of surface tension and viscosity. In Sections~\ref{sec:bips},~\ref{sec:ocps}~and~\ref{sec:tcps} we present our results
for self-assembly in critical and off-critical binary immiscible and
ternary amphiphilic fluids respectively. We close the paper with
discussion and conclusions in Section~\ref{sec:concs}.

\section{The lattice gas model}\label{sec:model}

Our lattice-gas model is based on a microscopic, bottom-up
approach, where dipolar amphipile particles are included alongside
the immiscible oil and water species.  Lattice-gas particles can have
velocities $\bfc_i$, where $1\leq i\leq b$, and $b$ is the number of
velocities per site.  We shall measure discrete time in units of one
lattice timestep, so that a particle emerging from a collision at site
$\bfx$ and time $t$ with velocity $\bfc_i$ will advect to site
$\bfx+\bfc_i$ where it may undergo the next collision.
We let $n^\alpha_i\bfxt\in\{0,1\}$ denote the presence ($1$) or absence
($0$) of a particle of species $\alpha\in\{R,B,A\}$ ($R$, $B$, $A$
denoting red (oil), blue (water) and green (amphiphile) species
respectively) with velocity $\bfc_i$, at lattice site $\bfx\in\calL$
and time step $t$. The collection of all $n^\alpha_i\bfxt$ for $1\leq
i\leq b$ will be called the {\it population state} of the site; it is
denoted by
\begin{equation}
\bfn\bfxt\in\calN
\end{equation}
where we have introduced the notation $\calN$ for the (finite) set of
all distinct population states.
The amphiphile particles also have an orientation denoted by
$\bfsigma_i\bfxt$.  This orientation vector, which has fixed magnitude
$\sigma$, specifies the orientation of the amphiphile particle at site $\bfx$ and time step $t$
with velocity $\bfc_i$. The collection of the $b$ vectors
$\bfsigma_i\bfxt$ at a given site $\bfx$ and time step $t$ is called
the {\it orientation state}. 
We also introduce the {\it colour charge} associated
with a given site,
\begin{equation}
q_i\bfxt\equiv n^R_i\bfxt-n^B_i\bfxt,
\end{equation}
as well as the total colour charge at a site,
\begin{equation}
q\bfxt\equiv\sum_{i=1}^b q_i\bfxt.
\end{equation}
The state of the model at site ${\bf x}$ and time
$t$ is completely specified by the
population state and orientation state of all the sites. 
The time evolution of the system is an alternation between an advection
or {\it propagation} step and a {\it collision step}.  In the first of
these, the particles move in the direction of their velocity vectors to
new lattice sites.  This is described mathematically by the replacements
\begin{equation}
n^\alpha_i\left(\bfx+\bfc_i,t+1\right)
\leftarrow
n^\alpha_i\bfxt,
\label{eq:propp}
\end{equation}
\begin{equation}
\bfsigma_i\left(\bfx+\bfc_i,t+1\right)
\leftarrow
\bfsigma_i\bfxt,
\label{eq:propo}
\end{equation}
for all $\bfx\in\calL$, $1\leq i\leq b$ and $\alpha\in\{ R,B,A\}$.  That
is, particles with velocity $\bfc_i$ simply move from point $\bfx$ to
point $\bfx+\bfc_i$ in one time step.
In the collision step, the newly arrived particles interact, resulting
in new momenta and surfactant orientations.  The collisional change in
the state at a lattice site $\bfx$ is required to conserve the mass of
each species present
\begin{equation}
\rho^\alpha\bfxt\equiv\sum_i^b n^\alpha_i\bfxt,
\end{equation}
as well as the $D$-dimensional momentum vector
\begin{equation}
\bfp\bfxt\equiv\sum_\alpha\sum_i^b\bfc_i n^\alpha_i\bfxt,
\end{equation}
(where we have assumed for simplicity that the particles all carry unit
mass).  Thus, the set $\calN$ of population states at each site is
partitioned into {\it equivalence classes} of population states having
the same values of these conserved quantities. 

We assume that the characteristic time for collisional and
orientational relaxation is sufficiently fast in comparison to that of
the propagation that we can model this probability density as the
Gibbsian equilibrium corresponding to a local, sitewise Hamiltonian function; that is
\begin{equation}
\calP(s')
=
\frac{1}{\calZ}\exp\left[-\beta H(s')\right], \label{eq:beta_defn}
\end{equation}
where $\beta$ is an inverse temperature, $H(s')$ is the energy
associated with collision outcome $s'$, and $\calZ$ is the
equivalence-class partition function.
The sitewise Hamiltonian function for  our model has been previously derived and
described in detail for the two-dimensional version of the
model~\cite{bib:bce}, and we use the same one here. It is
\begin{equation}
H(s') =
\bfJ\cdot (\alpha\bfE+\mu\bfP) +
\bfsigma'\cdot (\epsilon\bfE+\zeta\bfP) +
\calJ : (\epsilon\calE+\zeta\calP)+{\delta \over 2}
{{{\bf v}({\bf x},t)}^{2}},
\label{eq:hamil}
\end{equation}
where we have introduced the {\it colour flux} vector of an outgoing
state,
\begin{equation}
\bfJ\bfxt\equiv\sum_{i=1}^b\bfc_i q'_i\bfxt,
\label{eq:cflux}
\end{equation}
the {\it total director} of a site,
\begin{equation}
\bfsigma\bfxt\equiv\sum_{i=1}^b\bfsigma_i\bfxt.
\end{equation}
the {\it dipolar flux} tensor of an outgoing state,
\begin{equation}
\calJ\bfxt\equiv\sum_{i=1}^b\bfc_i\bfsigma'_i\bfxt,
\end{equation}
the {\it colour field} vector,
\begin{equation}
\bfE\bfxt\equiv\sum_{i=1}^b\bfc_i q\bfxtp,
\label{eq:bfEdef}
\end{equation}
the {\it dipolar field} vector,
\begin{equation}
\bfP\bfxt\equiv-\sum_{i=1}^b\bfc_i S\bfxtp,
\end{equation}
the {\it colour field gradient} tensor,
\begin{equation}
\calE\bfxt\equiv\sum_{i=1}^b\bfc_i\bfE\bfxtp,
\end{equation}
the {\it dipolar field gradient} tensor,
\begin{equation}
\calP\bfxt\equiv-\sum_{i=1}^b\bfc_i\bfc_i S\bfxtp,
\label{eq:calPdef}
\end{equation}
defined in terms of the scalar director field,
\begin{equation}
S\bfxt\equiv\sum_{i=1}^b\bfc_i\cdot\bfsigma_i\bfxt.
\label{eq:sdef}
\end{equation}
and the kinetic energy of the particles at a site,
\begin{equation}
{\delta \over 2}\left|{\bf v}({\bf x},t)\right|^2,
\end{equation}
where ${\bf v}$ is the average velocity of all particles at a site, the
mass of the particles is taken as unity, and $\alpha$, $\mu$,
$\epsilon$, $\zeta$ and $\delta$ are coupling constants. To maintain
consistency with previous work we use the coupling constants as previously
defined in  ~\cite{bib:bcp}. The values of these constants
are:
\begin{center}
$\alpha=1.0$,  $\epsilon=2.0$,   $\mu=0.75$,   $\zeta=0.5 $
\end{center}
These values were chosen in order to maximise the desired behaviour of
sending surfactant to oil-water interfaces while retaining the
necessary micellar binary water-surfactant phases. 
It should be noted that the inverse temperature-like
parameter $\beta$ (whose numerical value is varied in this paper) is not related in the conventional way to the kinetic
energy. For a discussion of the introduction of this parameter into
lattice gases we refer the reader to the
original work by Chen, Chen, Doolen and Lee~\cite{bib:2chendoolen}, and Chan and Liang~\cite{bib:chli}.
Eqs.~(\ref{eq:hamil})-(\ref{eq:calPdef}) were derived by assuming that
there is an interaction potential between colour charges, and that the
surfactant particles are like ``colour dipoles'' in this context~\cite{bib:bce}.
The term parameterised by $\alpha$ models the interaction of colour charges
with surrounding colour charges as in the original Rothman-Keller
model~\cite{bib:rk}; that parameterised by $\mu$ describes the interaction of
colour charges with surrounding colour dipoles; that parameterised by
$\epsilon$ accounts for the interaction of colour dipoles with surrounding
colour charges (alignment of surfactant molecules across oil-water
interfaces); and finally that parameterised by $\zeta$ describes the interaction of colour
dipoles with surrounding colour dipoles (corresponding to interfacial bending energy or
``stiffness''). This model has
been extensively studied in two dimensions
~\cite{bib:bce,bib:em1,bib:em2,bib:em3,bib:em4}, and the three-dimensional implementation employed in the present paper is described in
more detail by Boghosian, Coveney and Love~\cite{bib:bcp}. 

\section{Viscosity and surface tension}\label{sec:measure}

In this section we present the results of simulations designed to
measure the values of the macroscopic physical parameters which
control domain growth in fluid systems in our model, according to the
top-down theories described in Section 1. As emphasised above, these parameters are
the surface tension $\sigma$, the viscosity $\eta$ and the density $\rho$.

\subsection{Viscosity measurements}

We measured the viscosity by analysing the decay of shear waves. We
performed simulations to observe the decay of shear waves with an
initial velocity profile of the form:
\begin{equation}\label{eqn:profile}
{\bf v}(x,t)= v_{0}(t)\cos(2\pi \frac x N_{x}){\bf e}_{z}
\end{equation}
where $N_{x}$ is the lattice size in the $x$ direction, $v_{0}(t)$ is
the amplitude of the shear wave at time $t$, and ${\bf e}_{z}$ is the
unit vector in the $z$ direction. 
Solving the Navier-Stokes equations for the time evolution of $v_{0}(t)$ gives:
\begin{equation}
v_{0}(t)=v_{0}(0)\exp(-\nu(\frac {2\pi} {N_{x}})^2 t).
\end{equation}
Where  $v_{0}(0)$ is the initial amplitude of the shear wave and $\nu$ is
the kinematic viscosity.
We therefore initialise the system with a velocity profile given by~(\ref{eqn:profile}) and observe the decay of the shear wave by
calculating:
\begin{equation}
V_{z}(x,t)=\frac {1} {N_{y}N_{z}}  \sum_{yz} v_z(x,y,z,t),
\end{equation}
where $\sum_{yz}$ indicates summation over all lattice sites in the
$yz$ plane.
This gives the mean $z$ component of velocity, but includes any net $z$
velocity the system may possess. We subtract this to obtain the
velocity due only to the shear wave,
\begin{equation}
\widetilde{V}(x,t)=V_{z}-\frac {1} {N_{x}} \sum_{x} V_{z}(x,t),
\end{equation}
and calculate $v_{0}(t)$ as the r.m.s. value of this quantity,
\begin{equation}
v_{0}(t)=\sqrt{\frac {1} {N_{x}} \sum_{x} [\widetilde{V}(x,t)]^2}.
\end{equation}
We performed simulations at $5$ different amplitudes of initial velocity
profile and obtained kinematic viscosity $\nu = 0.78 \pm 0.01$ in
lattice units.

\subsection{Surface tension analysis}\label{sec:st}

In the present section we analyse the surface tension behaviour in our
model for both binary immiscible and ternary amphiphilic systems. The
central feature of ternary amphiphilic fluids is the lowering of the
interfacial tension between oil and water by the adsorption of
surfactant at the interface.
The existence of a spinodal point in a binary immiscible fluid is an
important feature of the fluid's thermal behaviour. At temperatures
above the spinodal point the fluid will not demix into single phase
domains. As one increases the temperature towards the spinodal point
the surface tension should be reduced to zero. 

We used a direct dynamical method for calculating the surface tension
across a flat interface.
In the vicinity of an interface the pressure is locally anisotropic,
as the pressure in the direction parallel to the interface is reduced
by the tension on the interface itself. For a system with flat
interface perpendicular to the $z$ axis the surface tension is given
by the line integral over $z$ of the component $P_{N}(z)$ of pressure
normal to the interface minus the component $P_{T}(z)$ tangential to
the interface~\cite{bib:mcc}: 
\begin{equation}\label{eq:anisigma}
\sigma=\int_{-\infty}^{\infty} \left( P_{N}(z)-P_{T}(z)\right) dz.
\end{equation}
Where:
\begin{equation}
P_{T}=P_{xx}=P_{yy}
\end{equation}
\begin{equation}
P_{N}=P_{zz}
\end{equation}

\begin{figure}[htp]
\centering
\includegraphics[width=0.8\textwidth,viewport=0 0 500 500,clip]{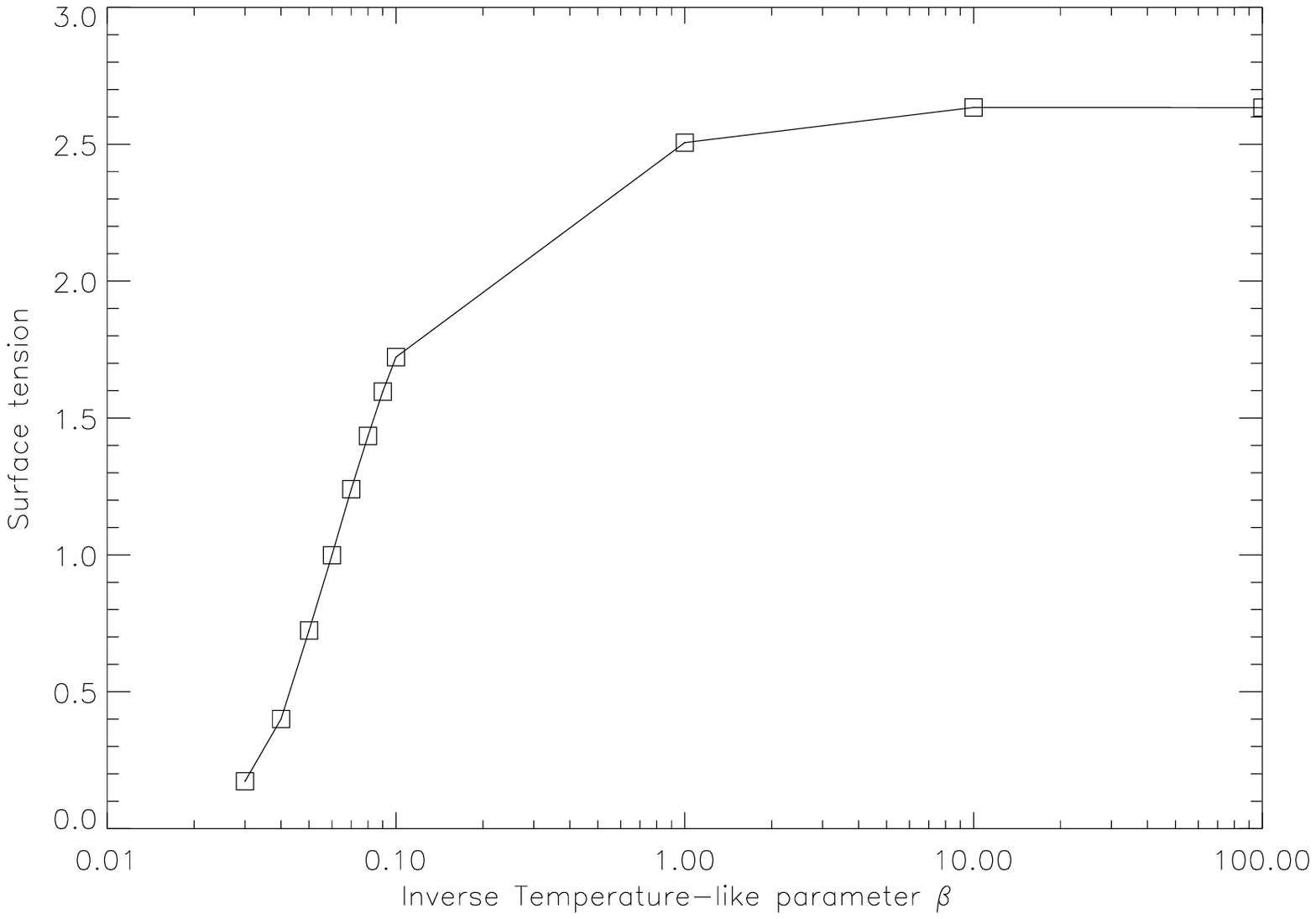}
\caption[fig:one]{Surface tension as a function  of
  inverse temperature-like parameter $\beta$ (both in lattice units)
  for a binary immiscible fluid. Error bars are
  not shown, but are smaller than the symbols. System sizes are $64^3$.}\label{fig:one}
\end{figure}

We begin by showing that our model is capable of reproducing
physically realistic interfacial tensions in systems of binary
immiscible fluids. We performed $12$ simulations for values of $0.02
\geq \beta \geq 100$, using systems of size $64^3$. The surface
tension was measured every time step for $1000$ time steps . The
simulation with $\beta=0.02$ was above the spinodal
point. The spinodal point itself was located by a computational steering
technique, which was found to be computationally more efficient than
traditional taskfarm methods. The value of
$\beta$ was modified during a simulation and the phase separation
behaviour visualised in real time. The value of $\beta$ at the
spinodal point was found to be $0.025 \pm 0.003$. This `compusteering'
technique represents a powerful and economic simulation technique, and
will be the subject of a future paper. Equilibration effects were found to be significant only close
to the spinodal point, and equilibration times for $\beta=0.03$,
$0.04$, $0.05$ were taken as $400$, $200$ and $200$ time steps
respectively. All other simulations were allowed to equilibrate for
$50$ time steps and the surface tension was then time averaged for the
remainder of the simulation. The surface tension as a function of
$\beta$ is shown in fig. (\ref{fig:one}).

We next analyse the behaviour of the surface tension as a function of
surfactant concentration adsorbed onto an oil-water interface. We use
two types of system, both initialised with a flat oil-water
interface perpendicular to the $z$ direction. The first system is
initialised with a fixed amount of surfactant at the interface. The second is initialised with the
surfactant dispersed throughout the system. We compute the surface
tension as above from eqn.~(\ref{eq:anisigma}).   
 
The equilibrium state for such a system in our model is quite complex. The
simplistic macroscopic picture of interfaces in ternary systems is
fairly static, with a monolayer of surfactant coating the interface
and lowering the surface tension. However, in reality as well as in
our model the surfactant may also exist in bulk solution far from the
interface, either as monomers or micelles or both. Equilibration of the
system is achieved when the rates for adsorption and desorption of monomers
and micelles are equal. These times are typically long, varying from
$10000$ to $20000$ time steps in our model. 

For the first type of system we simulate, where all surfactant particles
initially reside at the interface, the surface tension prior to
equilibration calculated by~(\ref{eq:anisigma}) is an increasing function
of time. The surfactant density at the interface decreases with time
as monomers and micelles desorb into the bulk. For the second type of
system, where all surfactant particles reside in the bulk, the
surfactant density at the interface increases as the surfactant adsorbs to the interface. The surface tension
calculated by~(\ref{eq:anisigma}) is a
decreasing function of time.

\begin{figure}[htp]
\centering
\includegraphics[width=0.8\textwidth,viewport=0 0 500 500,clip]{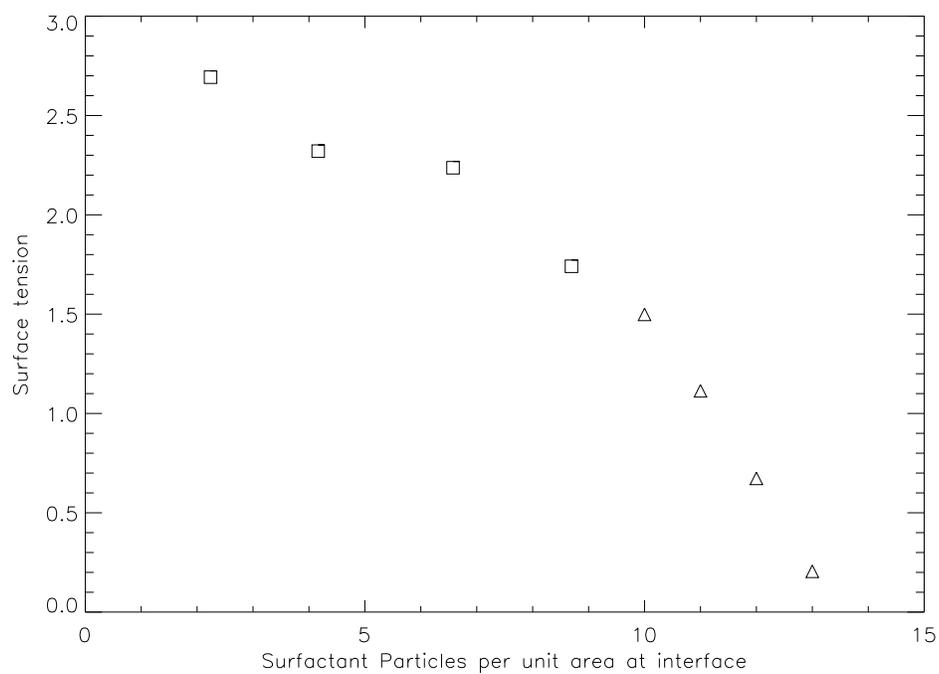}
\caption[fig:two]{Surface tension (lattice units) as a function  of
  surfactant concentration in a ternary system. Squares are values calculated after equilibration,
  triangles are calculated as a function of instantaneous surfactant
  density. All simulations have $\beta=1.0$ and were run on $64 \times
  32 \times 32$ lattices. }\label{fig:two}
\end{figure}

We calculate the surface tension as a function of surfactant density
in two ways. For low surfactant concentrations where we can reach
timescales on which the system is equilibrated we use the surface
tension and interfacial density for the equilibrated system. For
higher surfactant densities where the equilibration times become
prohibitive we plot the surface tension at time $t$ against the density
at time $t$. The results are shown in fig. (\ref{fig:two}).

\section{Critical Spinodal Decomposition}\label{sec:bips}

\subsection{Defining the characteristic size}\label{sec:corrfunct}

To analyse the domain growth quantitatively in the following
simulations we obtain the first zero crossing of the
coordinate-space pair correlation function. This defines the
characteristic domain size $R(t)$. The pair correlation
function is defined by:
\begin{equation}
C({\bf r}, t)=\frac{\int q({\bf x},t) q({\bf x}+{\bf r},t) d{{\bf x}}^{3}}{\int d{{\bf x}}^{3}}
\end{equation}
where $q({\bf x},t)$ is the colour charge at site
${\bf x}$, and the integral is taken over the whole system. 
We compute $C({\bf r}, t)$ for each of our simulations, and
perform an average over an ensemble of initial conditions. Taking the
spherical average of $C({\bf r}, t)$ yields $C(r,t)$, the first
zero of which gives the characteristic domain size. We obtain the
first zero by performing a linear interpolation between the last point
greater than zero and the first point less than zero. 

\subsection{Scaling analysis}
\begin{figure}[htp]
\centering
\includegraphics[width=0.8\textwidth,viewport=0 0 500 400,clip]{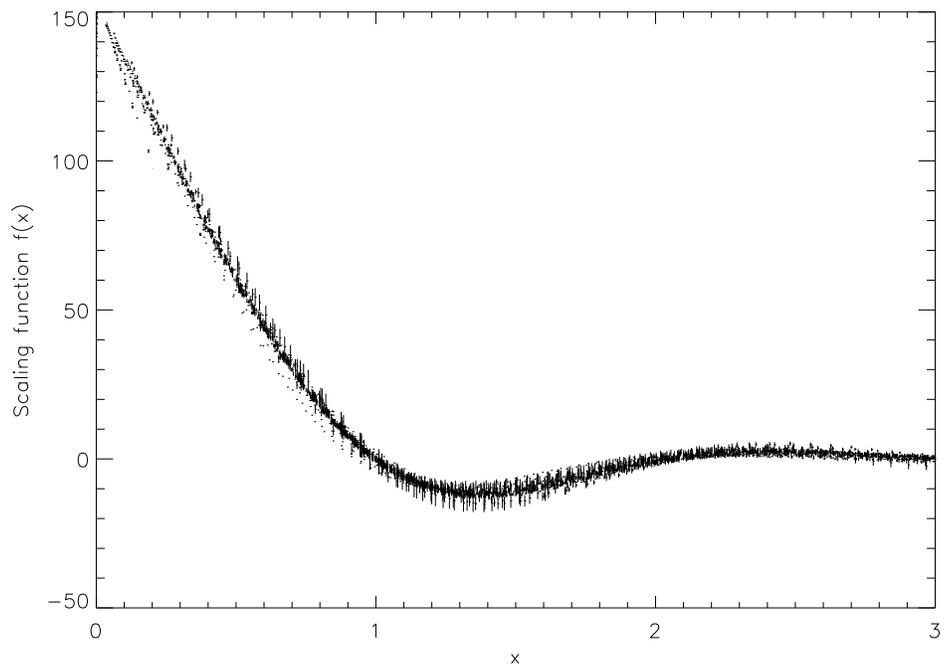}
\caption[fig:three]{The scaling function $f(x)$ as defined in eqn~(\ref{eq:scalfunct}) from the pair correlation function in critical binary phase
  separation for $0.08 \leq \beta \leq 100$ (lattice units). Data is taken from all
  simulations in this range of inverse temperature-like parameter $\beta$. System
  sizes are $64^3$ and $128^3$; data is taken between $100$ and $2300$
  time steps.}\label{fig:three}
\end{figure}

\begin{figure}[htp]
\centering
\includegraphics[width=0.8\textwidth,viewport=0 0 500 400,clip]{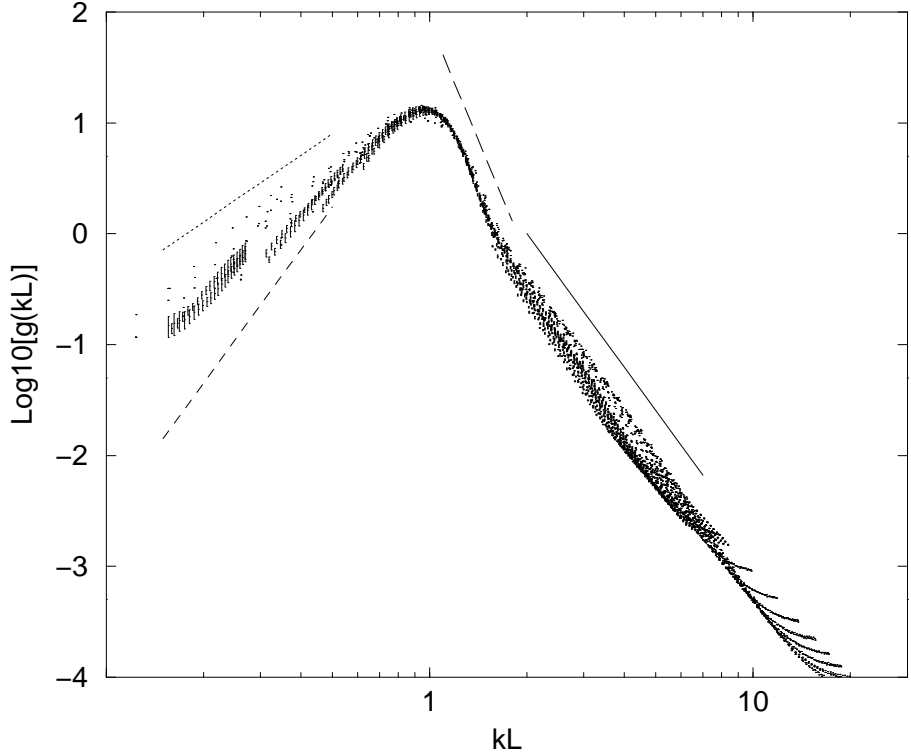}
\caption[fig:three]{The scaling function $g(x)$ as defined in eqn~(\ref{eq:scalfunct2}) from the structure factor in critical binary phase
  separation for $0.08 \leq \beta \leq 100$ (lattice units). Data is taken from all
  simulations in this range of inverse temperature-like parameter $\beta$. System
  sizes are $64^3$ and $128^3$; data is taken between $100$ and $2300$
  time steps. The dotted line is $(kL^2)$, the long dashed line is $(kL^4)$, the short dashed line is $(kL^{-7})$ and the solid line is $(kL^{-4})$. They are included as guides to the eye.}\label{fig:threeb}
\end{figure}

The first test of scaling is applied to the correlation functions of Section~\ref{sec:corrfunct}. If the domain structures are self-similar at different times during the coarsening, the scaling function $f(x)$ defined as
\begin{equation}\label{eq:scalfunct} 
C(r,t)=f{\Big(}\frac{r}{R(t)}{\Big)}=f(x)
\end{equation}
should be independent of $t$. Alternatively the scaling criteria may be applied to the Fourier transform of $C({\bf r}, t)$, the structure factor:  
\begin{equation}
S({\bf k}, t)=\frac{1}{N}{{ \Big \arrowvert} \frac{\int q({\bf x},t) \exp{(-i{\bf k}\cdot{\bf x}})d{{\bf x}}^{3} }{\int d{{\bf x}}^{3}}{\Big \arrowvert}}^2
\end{equation}
Spherically averaging this quantity yields S(k,t), which has a similar scaling form:
\begin{equation}\label{eq:scalfunct2} 
S(k,t)=L^{-3}g(kL)=L^{-3}g(q),
\end{equation}
where $g(q)$ is the fourier transform of $f(x)$. The functions $f(x)$ and $g(x)$ for simulations with $\beta >
0.07$ are plotted as a function of $x$ and $q$ respectively in figs. \ref{fig:three} and \ref{fig:threeb}. These datasets scale quite well, whereas
datasets from simulations with $\beta \leq 0.07$ do not. The specific form of g(q) has been the subject of much theoretical attention. The three features which receive the most attention are the small, intermediate and large  $q$ regimes. We show in figure \ref{fig:threeb} the behaviour of our model in these three regimes. In the large $q$ region we see a clear Porod tail with $g(q)\propto q^{-4}$. This region ends when $q$ probes the scale of the interface width. Furukawa speculated that there would be a regime $g(q) \propto q^{-6}$: we see a behaviour closer to $g(q) \propto q^{-7}$, in agreement with Jury et. al.~\cite{bib:jury}. In the small-q limit we see $g(q)\propto q^{-\delta}$, with $ 2 \le \delta \le 4$. In~\cite{bib:yeung} the constraint $\delta \le 4$ was derived for a Cahn-Hilliard theory without hydrodynamics. However, this derivation assumed a dynamical scaling exponent of $1/3$ and did not include fluctuations. Furukawa speculated that fluctuations would cause the $g(q) \propto q^{2} - g(q) \propto q^4$ crossover at small $q$.

\begin{figure}[htp]
\centering
\includegraphics[width=0.8\textwidth,viewport=0 0 500 400,clip]{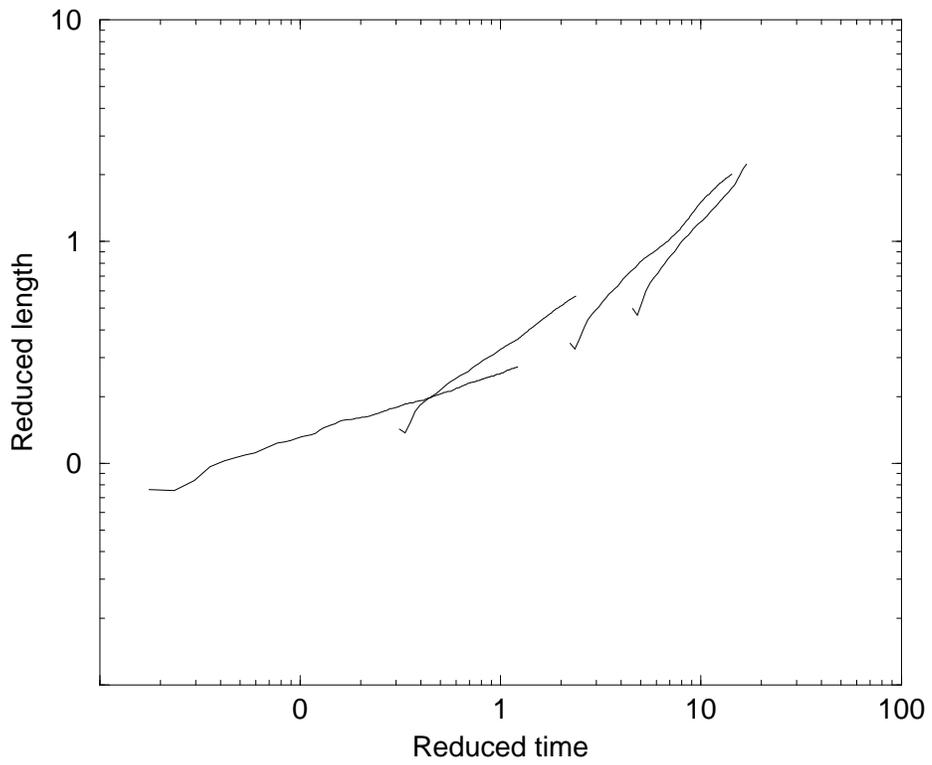}
\caption[fig:four]{Scaling plot in reduced variables $(l,t)$ for
  critical binary immiscible fluid phase separation. Data are from 
  simulations with  $\beta = 0.03, 0.04, 0.06, 0.08$ (from left to
  right). System sizes are $64^3$.}\label{fig:four}
\end{figure}

\begin{figure}[htp]
\centering
\includegraphics[width=0.8\textwidth,viewport=0 0 500 400,clip]{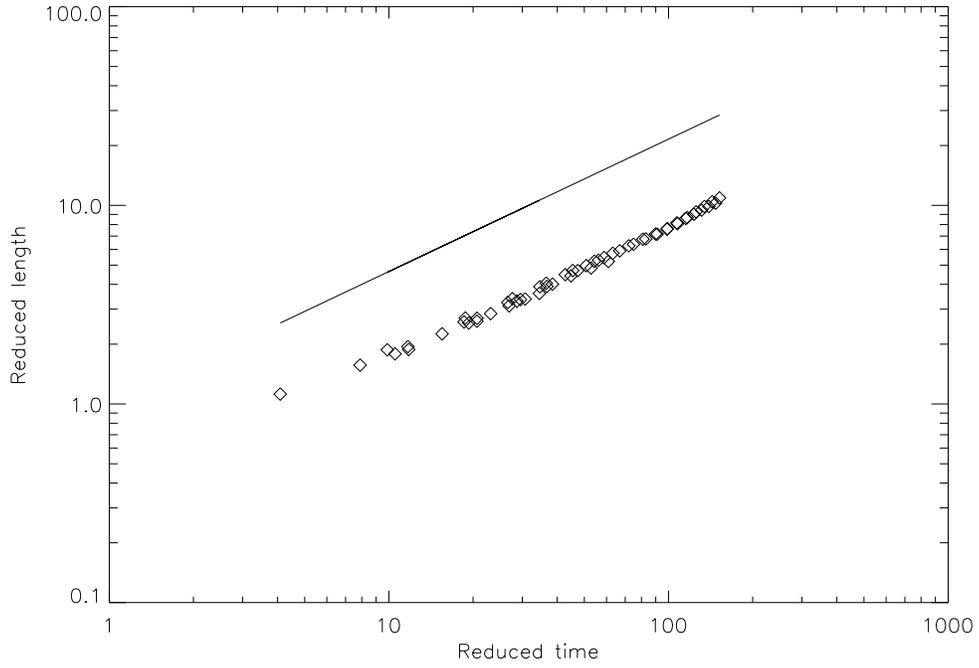}
\caption[fig:five]{Scaling plot in reduced variables $(l,t)$ for
  critical binary immiscible fluid phase separation. Data are from all
  simulations with  $0.10 \leq \beta \leq 100$. System
  sizes are $64^3$ and $128^3$; data is taken between $100$ and $2300$
  time steps. The solid line has gradient $2/3$ and is included as a
  guide to the eye only.}\label{fig:five}
\end{figure}

The above analysis of the correlation functions and structure factors is independent of the
form of $R(t)$. We now introduce the reduced length and time variables $l$ and
$t$ as defined in equations~(\ref{eq:red_len_defn})
and~(\ref{eq:red_time_defn}). To exclude finite
size effects we use data for which the characteristic domain size is
less than $\frac{1}{4}$ of the system size. Data from our simulations
satisfying this constraint spans a range of $ 3 <t< 152 $ and $1 <l<
10$.  A study using the DPD algorithm reached a range of $ 1<l<10^3$ and $1<t<5\times 10^4$, and a lattice-Boltzmann algorithm
has been used to reach a range of $ 1<l<2 \times 10^5$ and $1<t<5 \times
10^7$ ~\cite{bib:jury,bib:kendon}. The analysis method used to plot
the data displayed in figures \ref{fig:four} and \ref{fig:five} from simulations with
different values of $L_0$ and $T_0$ is identical to that of Kendon
{\it et al.}~\cite{bib:kendon}.

We find that data with $0.10 \leq \beta \leq 100$ show growth with an
inertial $(l \propto t^{2/3})$
exponent, and $R(t)$ collapses well onto a single scaling
function. Simulations with $\beta=0.03$ show slow growth where
$l \propto t^{1/3}$. Visualisation of the order parameter for this data
shows that sharp interfaces do not form during the simulation, and so data for these low values of surface tension do not satisfy the criteria for the postulated  universal scaling 
regime. This is confirmed by our scaling analysis of the correlation
functions above, where data for $ 0.03 \geq \beta \geq 0.07$ did not
collapse onto the same curve shown in figure 3. The time evolution
of the domain size in reduced units for $ 0.03 \geq \beta \geq 0.08$
is shown in figure \ref{fig:four}.

This inertial exponent is not consistent with the previous
results of Kendon {\it et al.} and Jury {\it et al.}~\cite{bib:jury,bib:kendon},
although it is similar to the results of Appert {\it et al.}~\cite{bib:appert}. In~\cite{bib:kendon}, the $(l=ct^{2/3})$
scaling of Appert {\it et al.} was ascribed to `excessive diffusion' in the
lattice gas algorithm. The particulate noise characteristic of the lattice gas is not present in their lattice-Boltzmann algorithm, and so it is perfectly plausible that these fluctuations continue to inhibit phase separation, reducing the growth exponent in the viscous regime to some effective exponent close to the $2/3$ usually associated with the inertial regime. An identical effect is well known in two dimensions, where models with fluctuations yield an exponent of $1/2$ in the early time regime, and models without yield an exponent of $1/3$~\cite{bib:wagneryeo,bib:em1}. 

This issue deserves some closer examination, however. If we interpret our structure factor and correlation function data as exhibiting goood scaling collapse (i.e. for all times considered the morphology is characterised by a single lengthscale), then the $g(q) \propto q^{2} - g(q) \propto q^4$ crossover at small q is consistent with speculations by Furukawa that fluctuations would cause the $q^2$ behaviour. As noted above these fluctuations may act to inhibit the phase separation and give rise to a lower effective exponent. 

Our immiscible lattice-gas model reduces to a simulation of a convection-diffusion equation for $\beta=0.0$. The diffusion constant in this equation is a function only of the collision rules. It may be possible by a judicious choice of collision rules or addition of rest particles to vary this diffusion constant independent of surface tension to fully investigate its effect.

If, however, we interpret the small $q$ behaviour as being indicative of poor scaling collapse, the exponent seen may be interpreted as a crossover between an early time and viscous regime. This issue could only be resolved by larger and longer simulations, and present hardware limitations mean that this must remain a matter for future investigation.

There is at present considerable uncertainty about both the universality of the asymptotic scaling and what the true asymptotic regime is. Theoretical work concerning the dispersion relation on fluid interfaces casts doubt on the universality of hydrodynamic phase separation in three dimensions~\cite{bib:yeung2}. 
Recent work by Kendon ~\cite{bib:kendonthesis} proposes a new set of macroscopically deduced scaling laws, based on the energy balance in the system. This analysis suggests that there may be more than one length scale of importance in spinodal decomposition. Another derivation of the growth laws proposes that the $n=\frac{2}{3}$ is transient and true asymptotic scaling is $n=\frac{4}{7}$~\cite{bib:solis}. However, it would be difficult to resolve such a subtle difference in exponent with any current numerical method. In addition, the derivation of~\cite{bib:solis} assumes a droplet morphology, even in the symmetric case. No such morphology is seen in any simulation of the symmetric case of which we are aware. The existence of multiple length scales and breakdown of scaling in two dimensional spinodal decomposition is well established~\cite{bib:wagneryeo,bib:wagnerscott}, and the existence of a single underlying scaling function in three dimensions remains an open question.

\section{Kinetics of phase separation for off-critical binary fluids }\label{sec:ocps}

In this section we analyse the dynamics of domain growth in systems
where the composition is asymmetrical. In such systems the domain
structure is not bicontinuous. The minority oil (or water) phase
exists as droplets, and the domains coarsen by diffusion of oil (or water)
from the bulk onto the droplet, and by droplet coalescence. The
composition $\phi$ for a system where oil is the minority phase is defined as:
\begin{equation}
\phi=\frac{\rho_o}{\rho_{o}+\rho_{w}},
\end{equation}
where $\rho_o$ and $\rho_{w}$ are the reduced densities of oil and
water respectively.
We performed simulations for values of $\phi=0.2$, and $\phi=0.4$. These
simulations were performed on $64^3$ systems for $3000$
time steps. The correlation function was averaged over five independent
simulations for $\phi=0.2$, and three independent simulations for
$\phi=0.4$. The results are shown in figures \ref{fig:seven} and \ref{fig:eight}.

\begin{figure}[htp]
\centering
\subfigure[]{\label{fig:sevena}\resizebox{0.4\textwidth}{!}{\includegraphics{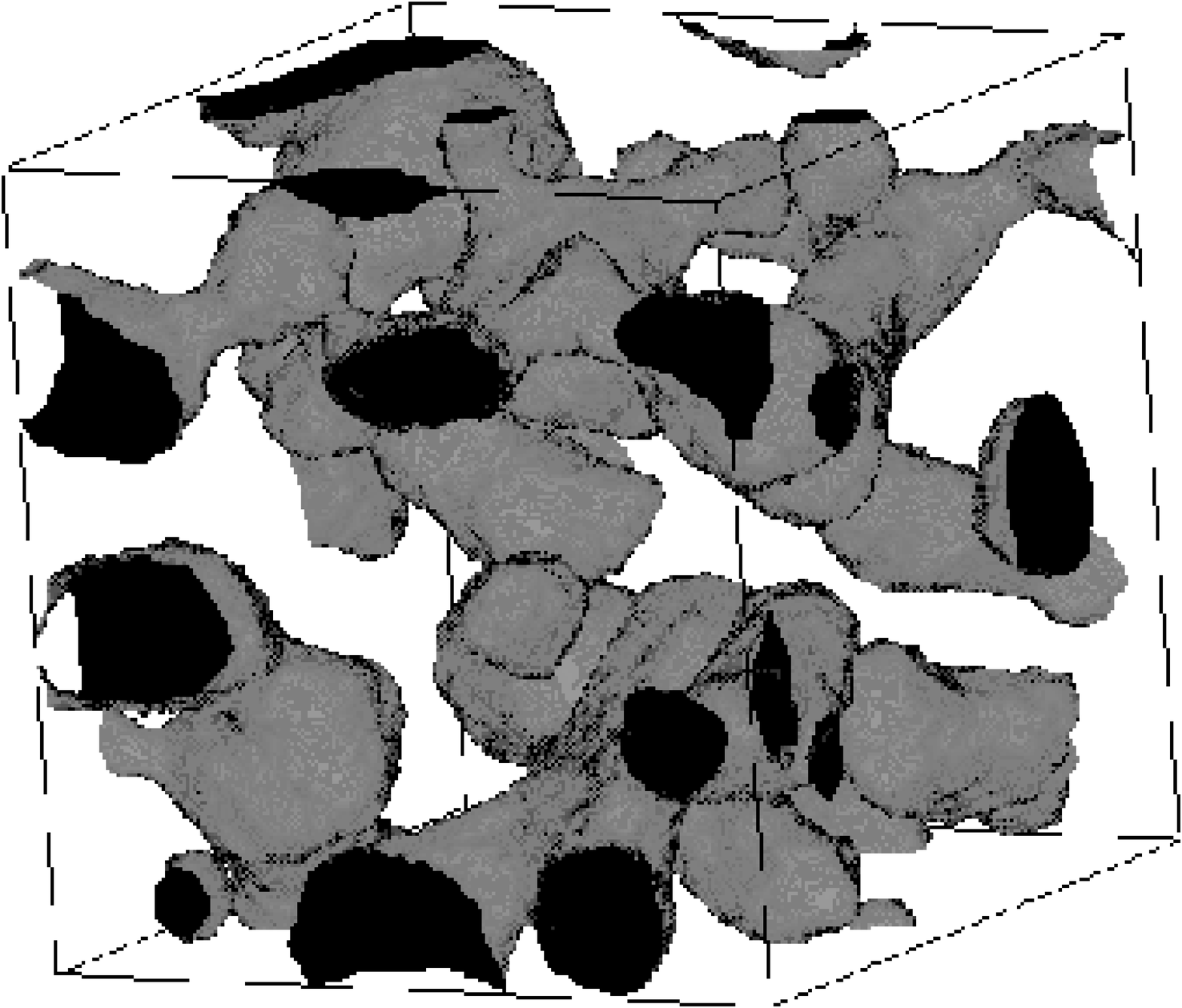}}}
\subfigure[]{\label{fig:sevenb}\resizebox{0.4\textwidth}{!}{\includegraphics{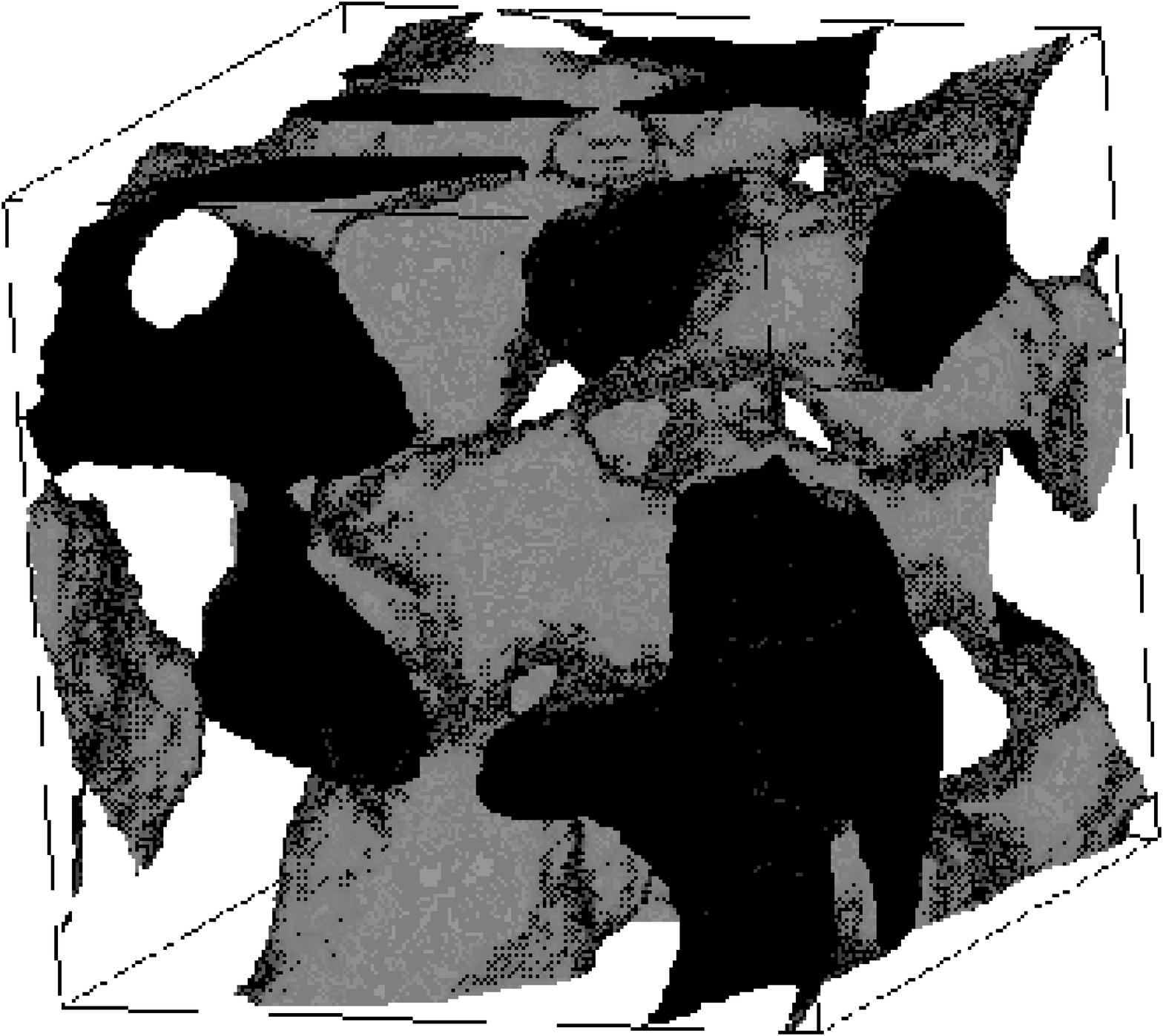}}}
\caption[fig:six]{Interface morphology at time step $500$ for minority
  phase (oil) volume fraction $0.20$ (a) and $0.40$ (b). System
  size is $64^3$.}\label{fig:six}
\end{figure}

\begin{figure}[htp]
\centering
\includegraphics[width=0.8\textwidth,viewport=0 0 500 400,clip]{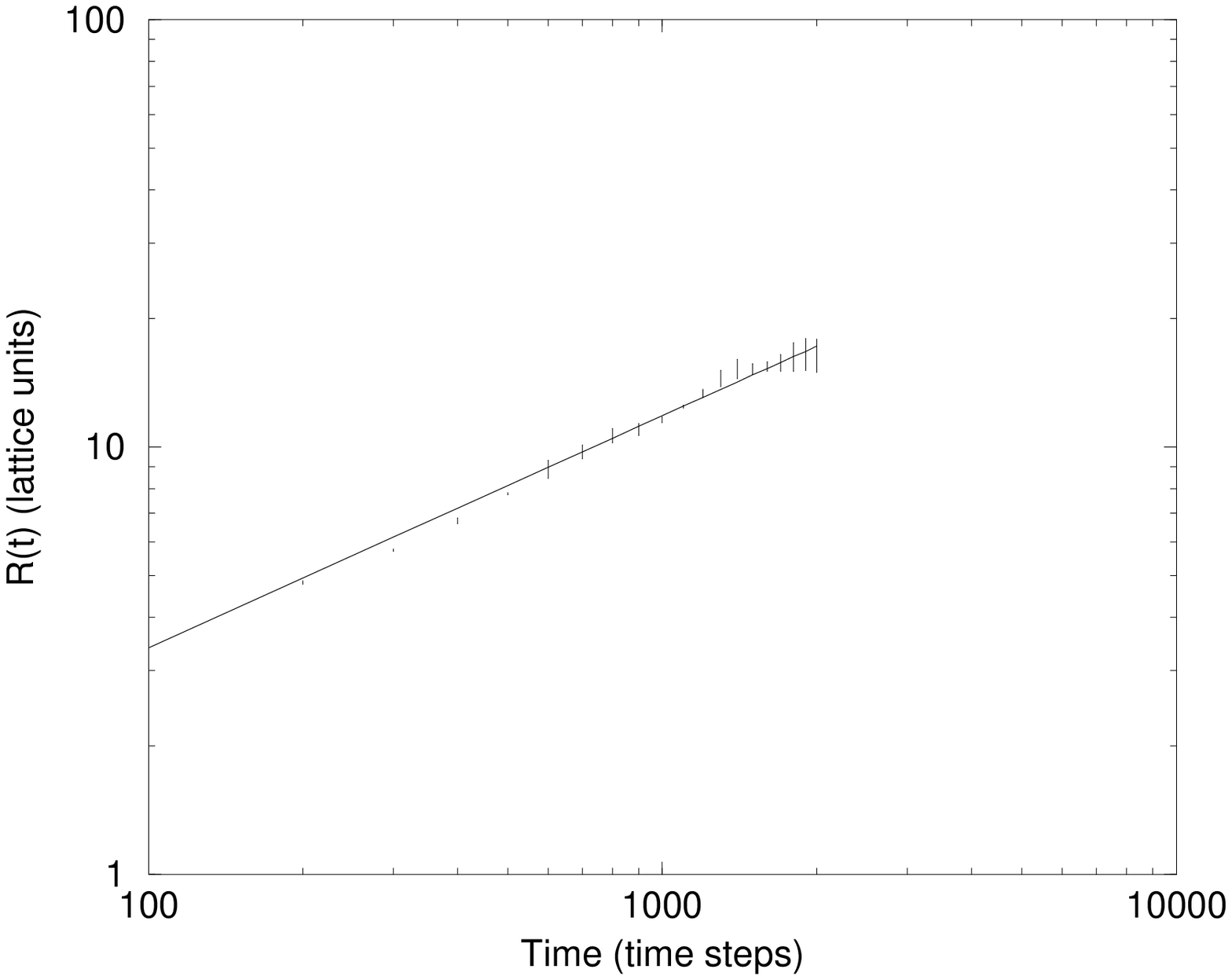}
\caption[fig:seven]{Scaling in off-critical binary phase separation for
  $\beta=1$ with composition $\phi=0.2$ (lattice units). The solid line is a least squares 
  fit with effective exponent $0.54 \pm 0.01$. Error bars show one
  standard deviation on the mean of five independent simulations.}\label{fig:seven}
\end{figure}

\begin{figure}[htp]
\centering
\includegraphics[width=0.8\textwidth,viewport=0 0 500 400,clip]{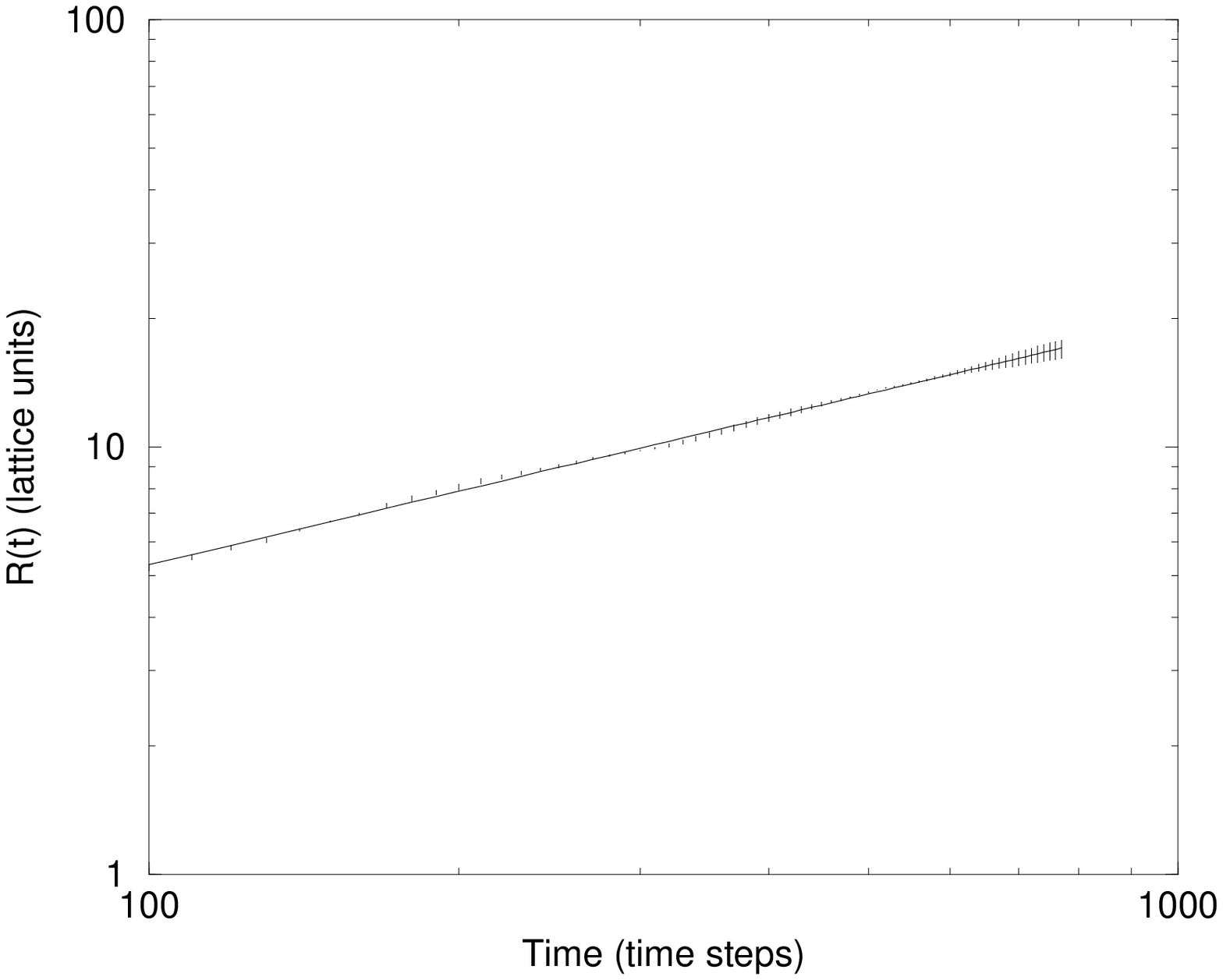}
\caption[fig:eight]{Scaling in off-critical binary phase separation for
  $\beta=1$, at a composition $\phi=0.4$ (lattice units). The solid line is a least squares 
  fit with effective exponent $0.573 \pm 0.003$. Error bars show one
  standard deviation on the mean of three independent simulations. }\label{fig:eight}
\end{figure}

The data in our simulations gives effective exponents of $0.543\pm 0.01$
and $0.573\pm0.003$ for compositions $0.2$ and $0.4$ respectively.
We do not observe the exponent $1/3$ expected from simple
theories of droplet coalescence and nucleation. However, as fig. \ref{fig:six} shows, the morphology for both these volume
fractions is far from being a collection of isolated drops. It is
likely that the exponent we measure is therefore intermediate between
the critical case and droplet coalescence and nucleation. The
existence of such intermediate exponents is well established. Previous
work by Appert {\it et al.}~\cite{bib:appert} on off-critical
decomposition for $\phi=0.05$ also saw growth more rapid than the
$t^{1/3}$ expected for simple nucleation and coalescence. The work
of~\cite{bib:weig} for the two-dimensional implementation of our
model similarly saw a continuous variation of exponent with
composition, as did~\cite{bib:novik1} for equal viscosity DPD fluids.

\section{Self assembly in ternary amphiphilic fluids }\label{sec:tcps}

\begin{figure}[htp]
\centering
%\subfigure[]{\label{fig:tena}\resizebox{0.3\textwidth}{!}{\includegraphics[1cm,2cm][12cm,13cm]{figure10a}}}
%\subfigure[]{\label{fig:tenb}\resizebox{0.3\textwidth}{!}{\includegraphics[1cm,2cm][12cm,13cm]{figure10b}}}
%\subfigure[]{\label{fig:tenc}\resizebox{0.3\textwidth}{!}{\includegraphics[1cm,2cm][12cm,13cm]{figure10c}}}
\subfigure[]{\label{fig:tena}\resizebox{0.3\textwidth}{!}{\includegraphics{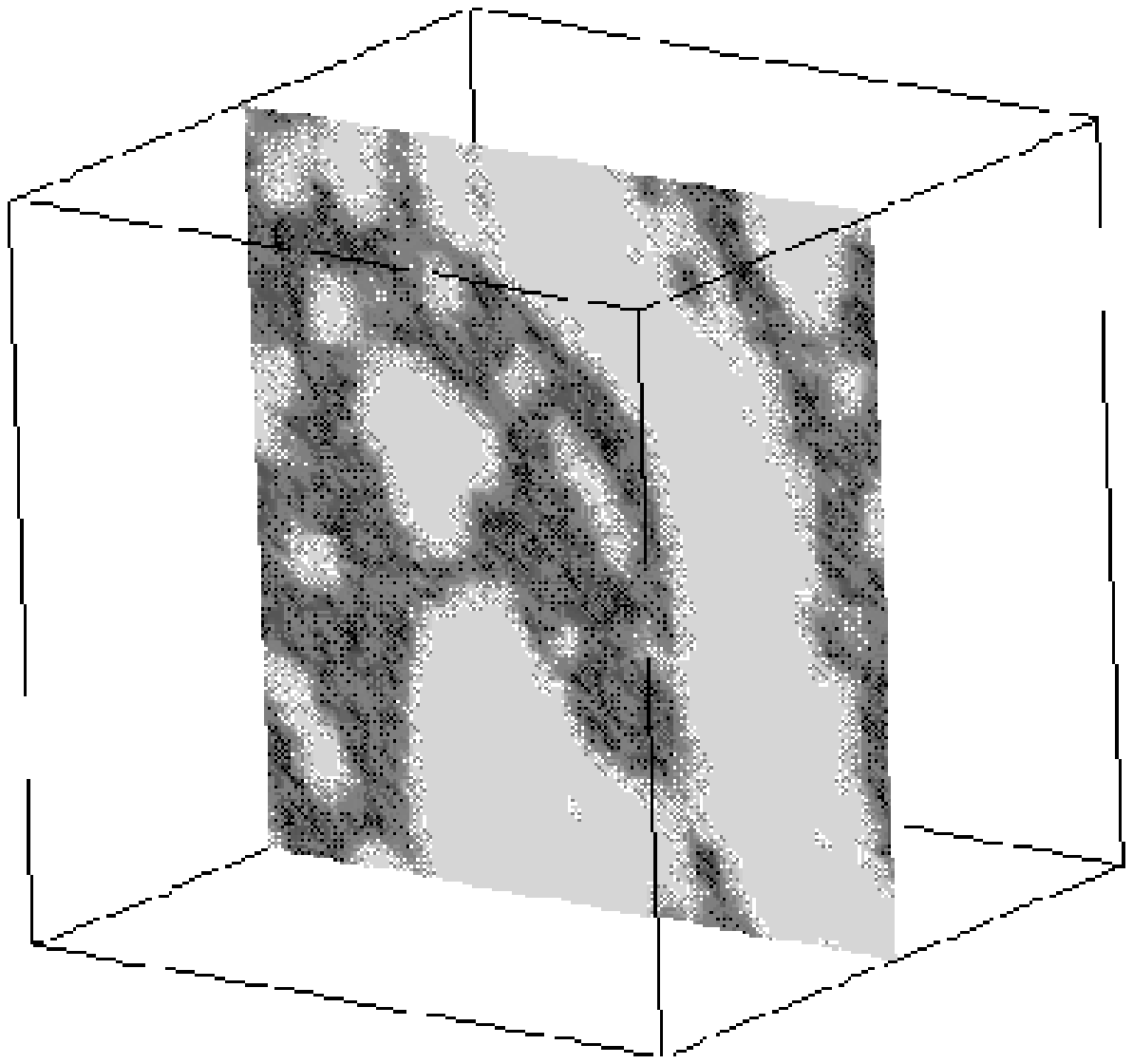}}}
\subfigure[]{\label{fig:tenb}\resizebox{0.3\textwidth}{!}{\includegraphics{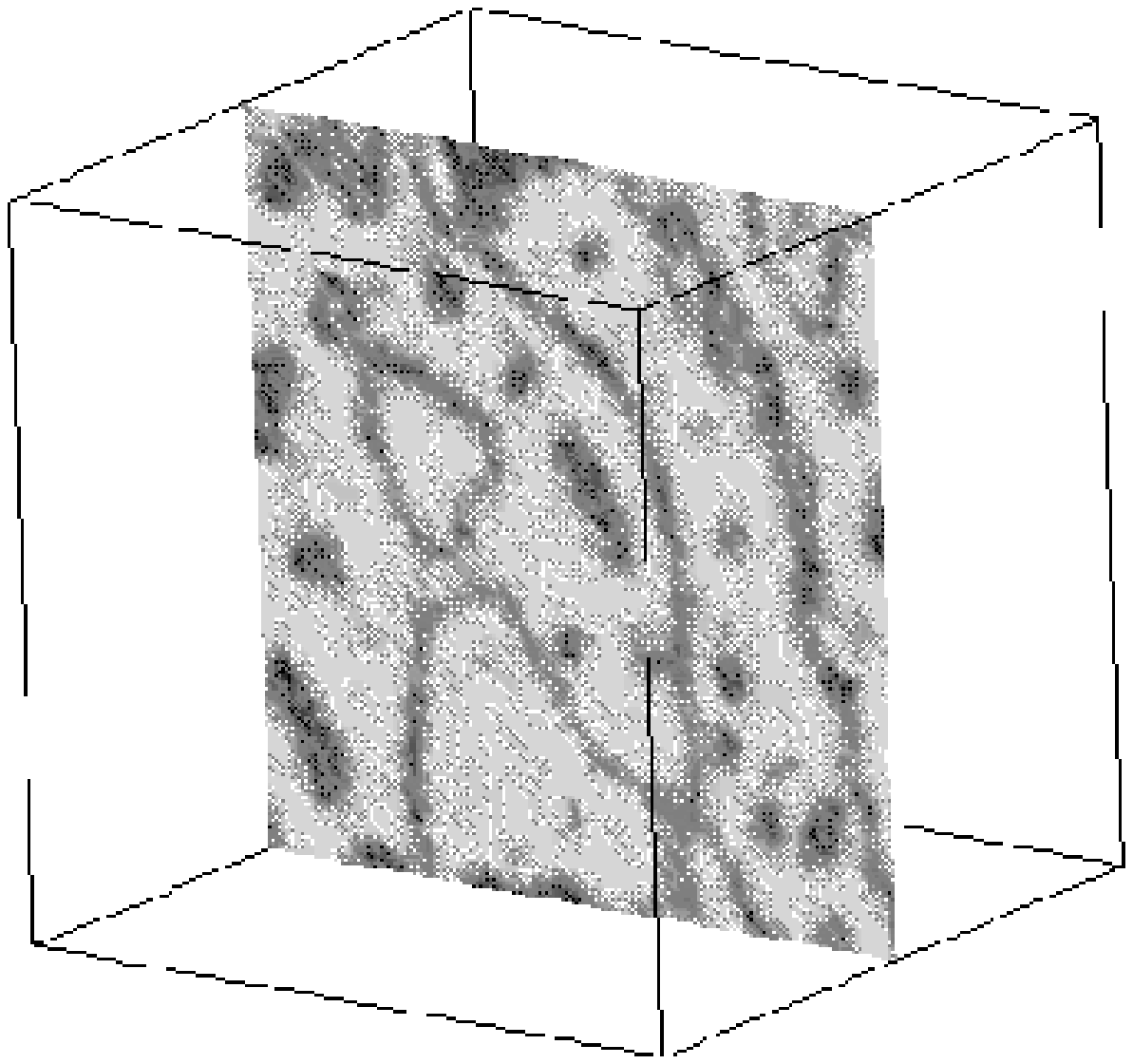}}}
\subfigure[]{\label{fig:tenc}\resizebox{0.3\textwidth}{!}{\includegraphics{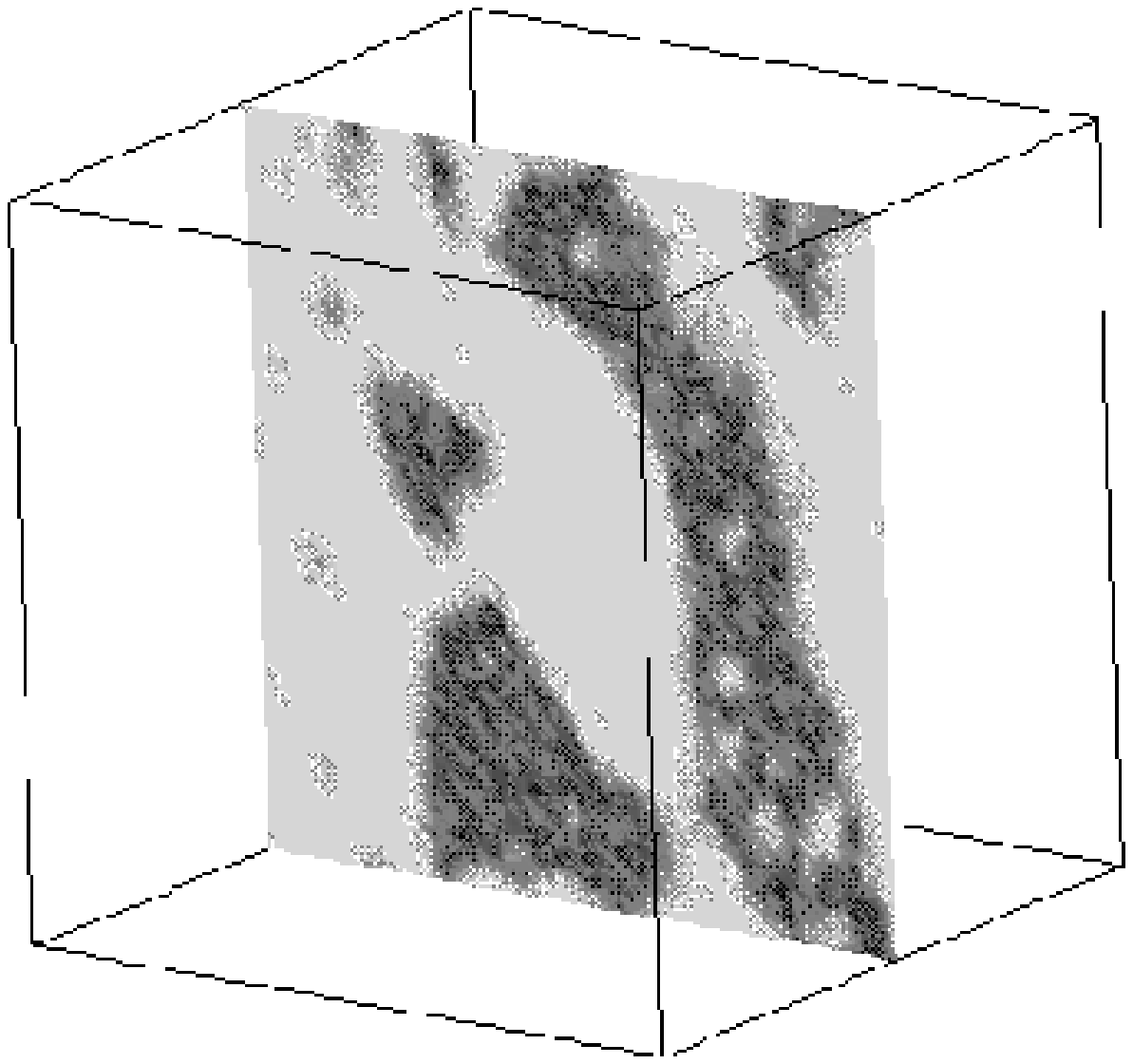}}}
\caption[fig:nine]{Sponge microemulsion phase at
  time step 850 following a random initialisation. (a) Water
  slice plane. (b) Surfactant slice plane, (c) Oil
  slice plane. The system size is $64^{3}$. Reduced densities of oil, water,
  and surfactant in this system are $0.19$,$0.19$ and $0.12$, respectively. }\label{fig:nine}
\end{figure}

In this section we turn to the analysis of ternary amphiphilic fluids. We
concentrate on systems with equal amounts of oil and water, varying
the amount of surfactant for each simulation. The presence of
surfactant dramatically alters the interfacial energetics and structure, and
consequently the dynamics of domain growth. Specifically, the
adsorption of surfactant at oil-water interfaces and its concomitant
lowering of the surface tension weaken the forces which drive
binary immiscible phase separation.  For sufficiently large
surfactant concentrations the final equilibrium state is a
bicontinuous or sponge {\it
  microemulsion} phase, where the domain growth is arrested at some final
characteristic domain size $R_c$. Such a system is depicted in
fig. (\ref{fig:nine}). All three components are bicontinuously
connected, with the surfactant particles lining the interface in a
monolayer between the percolating oil and water regions.

We work with $\beta=1.0$ and use the values for the coupling
coefficients in eqn~(\ref{eq:hamil}) as defined earlier 
(in Sec. ~\ref{sec:model}). The results that follow for the ternary
emulsion system have been obtained from a $64^3$ system with periodic boundary
conditions in all directions, the system having been
intialised from a random quench with the 
particles placed randomly on the lattice. The total reduced density of
the system was kept constant at $0.5$. The characteristic domain size
$R(t)$ was measured from the spherically averaged pair correlation function as
described in Section~\ref{sec:corrfunct}.
The growth of sponge as opposed to droplet phases in
these systems means that for the lowest values of the surfactant
concentration the growth behaviour should represent a perturbation of
the critical quenches investigated in Section~\ref{sec:bips}.
The results for reduced surfactant concentrations $0.02$, $0.06$, and $0.08$ support this conclusion.  In all cases the 
late-time behaviour is consistent with an algebraic growth law of the form
$R \sim t^{2/3}$. Prior to this there is an an early-time regime
in which the growth is consistent with a diffusive algebraic exponent
of $\frac{1}{3}$. Visualisation of the surfactant densities during this period
shows that surfactant adsorbs at the oil-water interfaces. 

Once the system reaches the biconinuous oil and water state
the inertial hydrodynamic regime begins. The length of the
diffusive period increases with increasing surfactant concentration
($50$ time steps for reduced surfactant concentration $\rho_{s}=0.02$, $60$ time steps
for $\rho_{s}=0.04$, $70$ time steps for $\rho_{s}=0.06$ and $100$
time steps for $\rho_{s}=0.08$). As one increases the surfactant
concentration the fluctuations in the domain size at late times
increases. This is consistent with the known dynamical nature of the
adsorption and desorption of surfactant to and from interfaces in ternary
systems as one approaches emulsification.  
With a reduced surfactant density of $0.10$ the time evolution of the
domain size ceases to follow the algebraic $R \sim t^{2/3}$ growth
law. We see a slower-than-algebraic growth law,  as previously observed
in the two dimensional implementation of our model where a growth
function $R(t)=(\ln t)^{\theta}$ was proposed,  based on a comparison with slow
growth in systems with quenched impurities
~\cite{bib:em1,bib:qd}. Consequently we look at a plot of $\ln t$ against
domain size in order to determine whether we have logarithmically slow
growth in this regime. The characteristic domain size $R(t)$ for
surfactant concentrations $0.10, 0.12, 0.14, 0.16$ are
plotted against $\ln t $ on logarithmic scales in fig. (\ref{fig:eleven}). One can
clearly see a transition through a regime where logarithmically slow
growth dominates throughout the timescale of the simulation. Such a
growth law is inconsistent with arrest of the domain growth at late
times for reduced surfactant density below $0.16$.
As we increase the reduced surfactant density beyond
$0.16$ we see complete arrest of the domain growth at late times.
We performed a fit of a stretched exponential function to datasets from
simulations with $\rho_{s}=0.18$. This function is defined identically
with~\cite{bib:em1}:
\begin{equation}
R(t)=A-B \exp(-Ct^D)
\end{equation}

These fits are shown in fig.~(\ref{fig:twelve}). To quantify
the effect that the
surfactant has on the domain size at late times in these simulations, following Emerton {\it et al.}~\cite{bib:em1} and Gyere
{\it et al.}~\cite{bib:qd}, we define $R_c$ as
  the domain size at time step $850$, where data which is unaffected by
  finite size effects is available for all surfactant concentrations. 
We plot $R_c$ against the inverse of the reduced surfactant density at the interface (having subtracted the micellar and monomeric concentration)
in fig. \ref{fig:thirteen} (for systems with quenched
impurities the amplitude of the disorder is analogous to the surfactant
concentration). We find a linear dependence of $R_c$ on $1/\rho_s$,
consistent with the results found in ~\cite{bib:em1} and ~\cite{bib:qd}.

\section{Conclusions}\label{sec:concs}

We have studied the dynamics of domain growth in binary immiscible fluids, in
both critical and off-critical cases, and in ternary
(oil-water-surfactant) emulsions and microemulsions,
using a three-dimensional hydrodynamic lattice gas model. In the
critical binary case we performed a scaling analysis similar to that
of Jury {\it et al.}~\cite{bib:jury} and Kendon {\it et al.}~\cite{bib:kendon}
and find late-time growth behaviour which is consistent with an
inertial hydrodynamic exponent. However the position of the late-time
domain growth on a plot of reduced time variables $(l,t)$ is
not consistent with that of ~\cite{bib:jury, bib:kendon}, in the same
way that the results of Appert {\it et al.}~\cite{bib:appert} showed an
inertial exponent in a crossover region of the reduced scaling plot. 
In the off-critical case we find effective exponents of $0.57$ and
$0.54$ for compositions $\phi=0.4$ and $0.2$ respectively. 
In the microemulsion case we observe a slowed algebraic growth
with exponent $\frac{2}{3}$, followed by a regime in which we see
evidence for logarithmically slow growth. For concentrations of
surfactant high enough to completely arrest domain growth we observe
good agreement with a stretched exponential fit to our data. Overall,
all our results are fully consistent with behaviour we observed previously 
in our two-dimensional lattice gas model.

\section*{Acknowledgements}

We are indebted to numerous people and organisations for their support
of and contributions to this work. They include Jean-Bernard Maillet,
Keir Novik, Nelido Gonzalez, Jeremy Martin, Mike Cates and Julia Yeomans. Simulations were performed on the Oscar
Origin 2000 service at Oxford Supercomputing Centre and the Cray T3E
at the Manchester CSAR service. Resources for the T3E were allocated
under EPSRC grant number GR/M56234. PJL
would like to thank EPSRC and Schlumberger Cambridge Research for
funding his CASE studentship award.  The collaboration between PVC and
BMB was supported by NATO grant number CRG950356.  BMB was supported
in part by the United States Air Force Office of Scientific Research
under grant number F49620-99-1-0070.

\begin{figure}[htp]
\centering
\includegraphics[width=0.8\textwidth,viewport=0 0 500 400,clip]{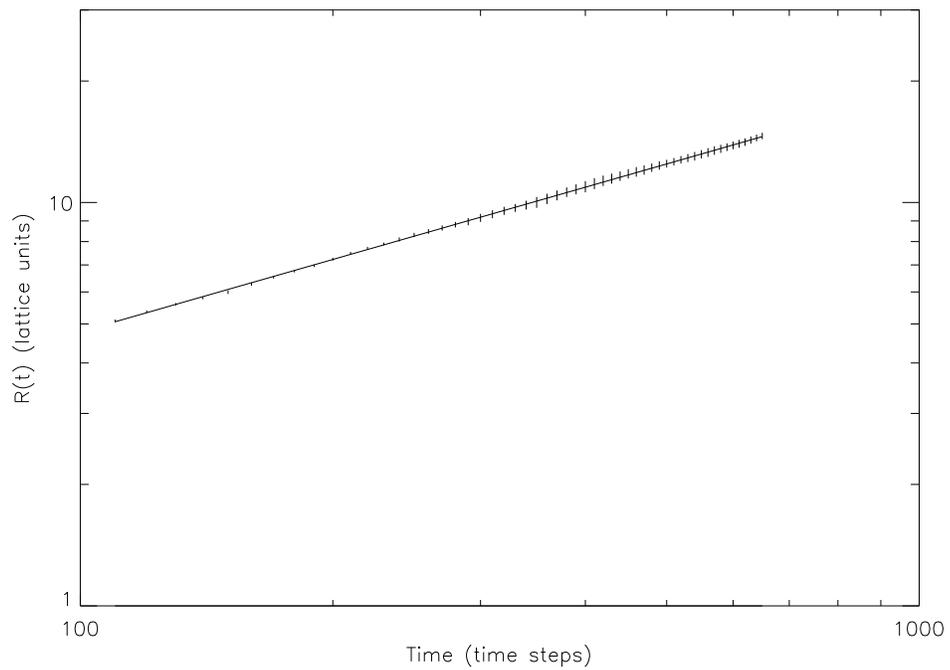}
\caption[fig:ten]{Time evolution of characteristic domain size
  (lattice units) for a
  ternary amphiphilic fluid with reduced densities of oil, water and
  surfactant $0.24$, $0.24$ and $0.02$ respectively. The solid
  line is a fit with an effective exponent of $0.59$. The error bars
  show one standard deviation on the mean of five independent
  simulations. The
  system size is $64^3$.}\label{fig:ten}
\end{figure}

\begin{figure}[htp]
\centering
\includegraphics[width=0.8\textwidth,viewport=0 0 500 400,clip]{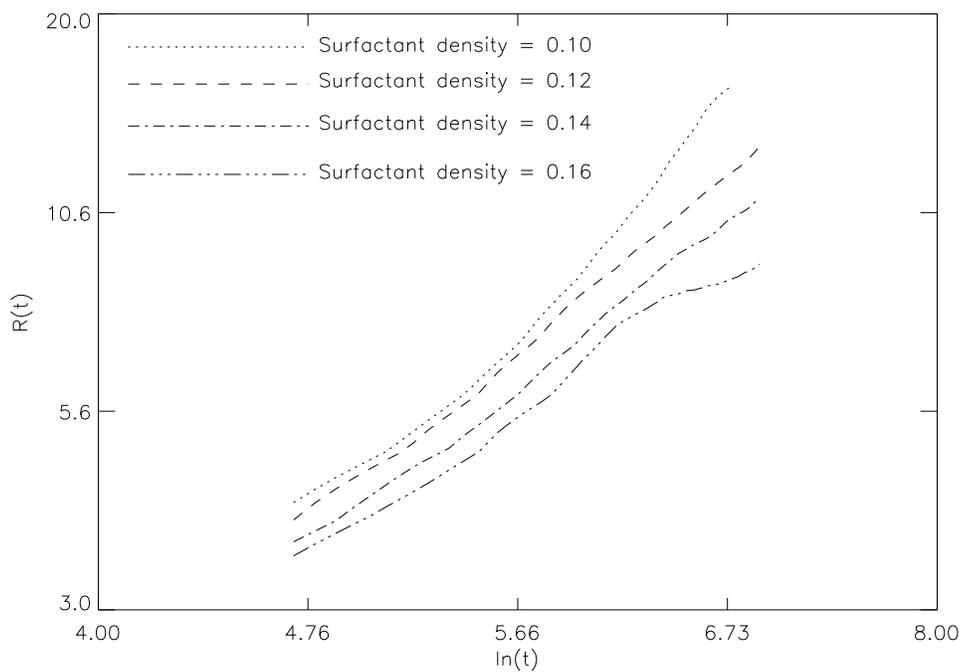}
\caption[fig:eleven]{ Crossover in ternary domain growth behaviour to logarithmically slow growth, and then to an intermediate regime. Surfactant
  concentration $0.10$ shows behaviour in a regime crossing over from
  algebraic to logarithmic growth. Surfactant concentrations
  0.12 and 0.14 show convincingly logarithmic growth, while
  concentration 0.16 shows time evolution in a regime between logarithmic and
  stretched exponential growth. Both time and characteristic size are
  measured in lattice units.   
}\label{fig:eleven}
\end{figure}

\begin{figure}[htp]
\centering
\includegraphics[width=0.8\textwidth,viewport=0 0 500 400,clip]{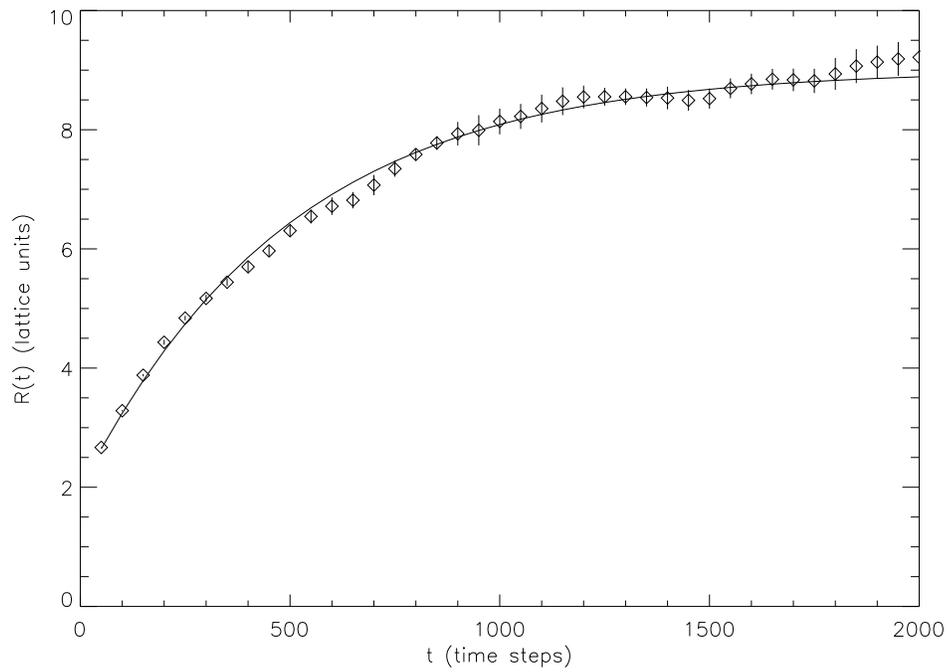}
\caption[fig:twelve]{Plot of characteristic
  domain size against time (both in lattice units) for a ternary system with reduced surfactant density
  $0.18$. The solid line is a least squares fit of a stretched
  exponential function to the data. Error bars show one standard
  deviation on the mean of five independent simulations.The system size is $64^3$.}\label{fig:twelve}
\end{figure}

\begin{figure}[htp]\label{fig:final_size}
\centering
\includegraphics[width=0.8\textwidth,viewport=0 0 500 400,clip]{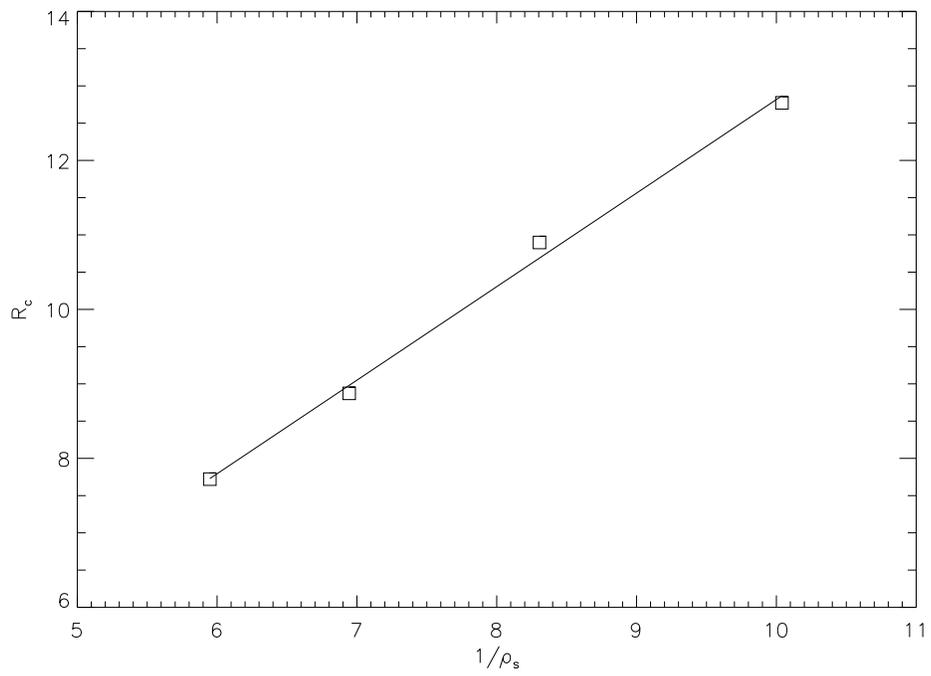}
\caption[fig:thirteen]{Plot of final characteristic
  domain size $R_c$ (lattice units) against the inverse of the reduced
  surfactant density of surfactant $1/\rho_s$ in the system. We have
  corrected $\rho_s$ by removing the micellar and monomeric surfactant
  concentrations.}\label{fig:thirteen}
\end{figure}

\end{document}